\documentstyle[osa,manuscript,psfig]{revtex}

\begin{document}

\titlepage

\title{Hyperon polarization in 
       $e^-p\to e^-HK$ with polarized electron beams}
\author{Liang Zuo-tang and Xu Qing-hua}
\address{Department of Physics, Shandong University,
Jinan, Shandong 250100, China}
\maketitle

\begin{abstract}
We apply the picture proposed in a recent Letter for 
transverse hyperon polarization in unpolarized 
hadron-hadron collisions to the exclusive process
$e^-p\to e^-HK$ such as
$e^-p\to e^-\Lambda K^+$,
$e^-p\to e^-\Sigma^+ K^0$,
or $e^-p\to e^-\Sigma^0 K^+$,
or the similar process $e^-p\to e^-n\pi^+$ 
with longitudinally polarized electron beams. 
We present the predictions for the longitudinal polarizations 
of the hyperons or neutron in these reactions, 
which can be used as further tests of the picture.
\end{abstract}

\newpage

\section{Introduction}

Hyperon polarization in different reactions have attracted 
much attention recently 
(see, e.g. [1-15], and the references given there). 
This is, on the one hand, because the polarization 
can easily be determined in experiments 
by measuring the angular distribution 
of their decay products thus provide us with a powerful tool 
to study the spin effects in different cases.
On the other hand, it is also triggered by  
the surprisingly large transverse hyperon polarization 
discovered already in 1970s in unpolarized 
hadron-hadron and hadron-nucleus reactions\cite{Bravar98}. 
Understanding the origin(s) of such puzzling transverse 
hyperon polarizations has been considered as a challenge 
to the theoretician working in this field and should 
provide us with useful information 
on spin structure of baryon and  
spin dependence of strong interaction. 

In a recent Letter\cite{Letter97}, triggered by the similarities of 
the corresponding data\cite{Bravar98,Asy00}, 
we pointed out that 
the transverse polarization observed\cite{Bravar98} in 
unpolarized hadron-hadron collisions should be 
closely related to the left-right asymmetries 
observed\cite{Bravar98,Asy00} in 
singly polarized $pp$ collisions.
They should have the same origin(s).
By using the single-spin left-right asymmetries for 
inclusive $\pi$ production as input, we can naturally 
understand the transverse polarization for hyperon 
which has one valence quark in common with the projectile, 
such as $\Sigma^-$, $\Xi^0$ or $\Xi^-$ in $pp$-collisions, 
or $\Lambda$ in $K^-p$-collisions.  
To further understand the puzzling transverse polarization 
of $\Lambda$ in $pp$-collisions, 
which has two valence-quarks in common with the projectile, 
we need to assume that the $s$ and $\bar s$, 
which combine respectively with 
the valence-$(ud)$-diquark and the remaining $u$-valence quark to 
form the produced $\Lambda$ and the associatively produced $K^+$ in 
the fragmentation region, should have opposite spins. 
Under this assumption, we obtained a good quantitative fit to 
the $x_F$-dependence of $\Lambda$ polarization in $pp$ collisions
(where $x_F\equiv 2p_\parallel/\sqrt{s}$, $p_\parallel$ is the 
longitudinal component of the momentum of the produced hyperon, 
and $\sqrt{s}$ is the total center of mass energy of the $pp$ system.) 
The obtained qualitative features for 
the polarizations of other hyperons are all 
in good agreement with the available data. 
We further pointed out, in a recent Brief Report\cite{BR00},
that $\Lambda$ polarization in the exclusive 
process $pp\to p\Lambda K^+$ can be used to give a special test 
of the picture since $|P_\Lambda|$ in this channel, which is 
the simplest one for $pp\to \Lambda X$, should take 
the maximum among all the different channels. 
It is encouraging to see that also this result 
is in agreement with the data 
obtained by R608 Collaboration at CERN\cite{R608}, 
which shows that $|P_\Lambda|$ in this 
channel is indeed much larger than that for $pp\to\Lambda X$.
It should be interesting to see whether 
the assumption that $s$ and ${\bar s}$ have opposite spins
can also be applied to other reactions.  
In this connection, it is encouraging that 
the measurements on $\Lambda$ polarization in 
$e^-p\to e^-\Lambda K^+$ is being carried out 
at Jefferson Laboratory\cite{Private}.

In this paper, we would like to apply the picture 
in [\ref{Letter97}] to $e^-p\to e^-\Lambda K^+$ 
to give the prediction on $\Lambda$ polarization 
in reactions with longitudinally polarized electron beams.
We compare the results with those obtained in other
cases for the spin states
of the $s$ and ${\bar s}$ needed for the production of 
$\Lambda$ and $K^+$ in this process. 
These will be given in Section 2. 
After that, in section 3, 
we extend the calculations to other similar 
processes such as $e^-p\to e^-\Sigma K$ or 
$e^-p\to e^-n\pi^+$ and 
present the predictions on the polarizations
of the hyperons (or neutron) in these reactions.  
We will also discuss the influence from the 
sea and that due to the mass differences 
between the $J^P=(1/2)^+$ and $J^P=(3/2)^+$ baryons 
in section 4. 
We will summarize the results in section 5. 
We found out that these 
polarizations are very sensitive to the 
spin states of the $s$ and $\bar s$ 
thus provide us with an
ideal tool to study the spin correlations between them. 

\section{Polarization of $\Lambda$ in $e^-p\to e^-\Lambda K^+$}

We now consider the process $e^-p\to e^-\Lambda K^+$, 
and discuss the problem in the center of mass frame 
of the produced hadronic system. 
We suppose that the energy and momentum transfer 
in the reaction 
are high enough so that the parton picture can be used. 
We thus envisaged with the following picture.
In the collision process, 
a $u_v$ quark is knocked out by the virtual photon.  
The knocked $u_v$ combines with a $\bar s_s$ to form the 
observed $K^+$ and the remaining $(u_vd_v)$-diquark 
combines with the $s_s$ 
to form the observed $\Lambda$. 
This process is illustrated pictorially in Fig.1. 

Since we are now considering the longitudinal polarizations 
of hyperons in $e^-p\to e^-HK$, 
two points of the picture in [\ref{Letter97}] apply:
(i) The $q_s$ and $\bar q_s$ that combine with the diquark 
from the incoming proton to form the hyperon and 
the remaining quark to form the associatively produced meson 
should have opposite spins. 
(ii) The SU(6) wave function can be used 
to relate the spins of the quarks to that of the hadrons. 
Using these two points, we obtain immediately the following:
Because $K^+$ is a spin zero object, the $\bar s_s$ has to have 
the spin opposite to the scattered $u_v$-quark.
Hence, according to (i), the $s_s$ quark should be polarized 
in the same direction as the scattered $u_v$-quark. 
According to the SU(6) wave function, 
which states that $|\Lambda^\uparrow \rangle =(ud)_{0,0}s^\uparrow$, 
[where the subscripts denote the spin and its third component of 
the $(ud)$-diquark], 
the $(ud)$-diquark has to be in the spin zero state
and the $\Lambda$ spin just equals to that of the $s$ quark.
Since the momentum of $\Lambda$ is opposite to that of $K^+$ or 
that of the scattered $u_v$-quark, 
the longitudinal polarization of $\Lambda$
is opposite to that of the scattered $u_v$-quark 
in the helicity basis.
Hence, we obtain,
\begin{equation}
P^{(a)}_\Lambda=-P_u=-P_eD(y).
\end{equation}
Here, $P_e$ is the polarization of the incoming electron beam. 
We use the superscript (a) to denote that  
the result is obtained in the case that the spins of
$s_s$ and $\bar s_s$ are opposite for the convenience of
comparison with other cases 
which will be discussed in the following.  
$D(y)$ is the spin transfer factor from $e^-$ to $q$ 
in the scattering process $e^-q\to e^-q$. 
This is an electromagnetic process thus $D(y)$ can easily 
be calculated and can be found in different publications. 
It is a function of only one variable $y$, 
which is defined as 
$y\equiv p\cdot q/p\cdot k$, where $p$, $k$ and $q$ 
are the four momenta for the incoming proton, 
$e^-$ and the virtual photon exchanged in the scattering. 
$D(y)$ is given as,
\begin{equation}
D(y)={1-(1-y)^2 \over 1+(1-y)^2}.
\end{equation}  
It is a monotonically increasing function of $y$ which increases 
from 0 to 1 when $y$ goes from 0 to 1.
Since $y$ is related to $x_B$ and $Q^2$ by $Q^2=sx_By$, 
(where $x_B$ is the Bjorken-$x$, $Q^2\equiv -q^2$, 
$s=2p\cdot k$ is the total $e^-p$ center of mass energy squared), 
it is clear that we can also express $P_\Lambda$ 
for $\Lambda$ in $e^-p\to e^-\Lambda K^+$ 
as a function of $x_B$ and $Q^2$ at given $\sqrt{s}$.

Eq.(1) is the result obtained based on the picture in Ref.[2]
under the assumption that the spins of
$s_s$ and $\bar s_s$ are opposite to each other.
To compare, we now consider the other two cases for 
the spins of $s_s$ and $\bar s_s$, i.e., 
(b) the spins of $s_s$ and $\bar s_s$ are parallel to each other,
and (c) the spins of $s_s$ and $\bar s_s$ are completely uncorrelated,
i.e., they have the same probability to be parallel or anti-parallel.
Apparently, in case (b), we have,
\begin{equation}
P^{(b)}_\Lambda=P_u=P_eD(y),
\end{equation}
and in case (c),
\begin{equation}
P^{(c)}_\Lambda=0.
\end{equation} 
We see that, the results are very much different from each other.
Hence, measuring $P_{\Lambda}$ in this process 
can distinguish these three different cases
with good efficiency.
             
\section{Polarization of $\Sigma$ or $n$ 
 in $e^-p\to e^-\Sigma K$ or $e^-p\to e^-n\pi^+$}
          
We can extend the calculations to other similar processes 
such as $e^-p\to e^-\Sigma^+K^0$,   
$e^-p\to e^-\Sigma^0K^+$,   
or $e^-p\to e^-n\pi^+$. 
As we can see from Fig.1, since in all these processes, 
the produced meson $K$ or $\pi$ 
is a spin zero object, we can obtain the same relation between 
the polarization of the $q_s$ shown in the figure and
that of the scattered $q_v$ from the incoming proton
as we obtained in $e^-p\to e^-\Lambda K^+$.
However, in contrast to the case for $\Lambda$ production 
where only spin zero $(ud)$-diquark contributes 
and if the $(ud)$-diquark is in the spin zero state, 
the produced $(ud)s$ can only be a $\Lambda$,  
in reactions such as $e^-p\to e^-\Sigma K$,
the diquarks in different spin states can contribute
and given the spin state of the diquark $(q_vq_v)$ 
and that of the quark $q_s$, the produced $(q_vq_v)q_s$ can still 
be a $J^P=(1/2)^+$ or a $J^P=(3/2)^+$ baryon.  
All these make the calculations more complicated. 
To obtain the polarization of the produced hyperon, 
we need to know the probabilities for the 
diquark which participates in the reaction 
to be in different spin states and the relative probabilities 
for the the produced $(q_vq_v)q_s$ to evolve into 
a $J^P=(1/2)^+$ or a $J^P=(3/2)^+$ baryon. 
The former can be calculated 
using the SU(6) wave-function of the proton. 
To do this, we note that, due to the helicity conservation, 
the scattered quark has the same longitudinal polarization 
before and after the scattering. 
Given the polarization of the scattered quark and the wave function 
of the proton, we can calculate the probabilities for 
the remaining diquark in different spin states. 
The latter can be determined 
by the projection of the spin wave function 
for the produced $(q_vq_v)q_s$ to that of the corresponding baryon, 
if we neglect the influence from the mass difference 
of the corresponding $(1/2)^+$ and $(3/2)^+$ baryons. 

To show how these calculations are carried out, 
we now take 
$e^-p\to e^-\Sigma^+K^0$ as an explicit example. 
In this reaction, the scattered quark has to be a $d$ quark. 
We recall that the SU(6) wave function for proton is,
\begin{equation}
|p^\uparrow\rangle ={1\over\sqrt{3}}[\sqrt{2}(uu)_{1,1}d^\downarrow
        -(uu)_{1,0}d^\uparrow],
\end{equation}
where we use the up or down arrow 
to denote the helicity ``$+$'' or ``$-$'' state.
We now consider the case where 
the scattered $d_v$ quark is 
in the helicity $+$ state. 
We see that if the incoming proton is in the helicity ``$+$'' state, 
the remaining $(u_vu_v)$-diquark 
should be in the $(u_vu_v)_{1,0}$ state, 
and the relative weight is $1/3$. 
The incoming proton can also be in the helicity ``$-$'' state. 
The corresponding wave function can be obtained by reversing 
the spins of the quarks in Eq.(5). 
In this case, 
since the scattered $d_v$ is in the helicity ``$+$'' state, 
the remaining $(u_vu_v)$-diquark should be 
in the $(u_vu_v)_{1,-1}$ state, 
and the relative weight is $2/3$. 
Hence, we obtain that, 
if the the scattered $d_v$ quark is in the helicity ``$+$'' state, 
the $(u_vu_v)$-diquark can be in  
$(uu)_{1,0}$ and $(uu)_{1,-1}$ state, 
and the relative probabilities are $1/3$ and $2/3$ respectively. 

Next, we consider that the produced 
$(u_vu_v)$-diquark combine together with a $s_s$ quark 
to form a $\Sigma^+$ or a $\Sigma^{*+}$. 
We first consider the case (a) that $s_s$ and ${\bar s}_s$
have opposite spins.
Since the $s_s$ should be polarized in the  
same direction as the scattered $d_v$ and
moves in the opposite direction of the scattered $d_v$, 
it should be in the helicity ``$-$'' state. 
We consider the case that the $(u_vu_v)$-diquark is in
the state $(u_vu_v)_{1,0}$. 
The spin state of the produced $(uus)$ is given by 
$(uu)_{1,0}s^\downarrow$. 
The projections of this state $(uu)_{1,0}s^\downarrow$ 
to the SU(6) wave function of $\Sigma$ and that of $\Sigma^*$ 
are $1/\sqrt{3}$ and $\sqrt{2/3}$ respectively. 
We thus obtain the relative probabilities for the production 
of these two hyperons in this case are $1/3$ and $2/3$ respectively. 
The results in other cases can be obtained in the same way. 
The results in case (c) are shown in Tables 1-3 for the 
different reactions.
Those in case (a) and (b) can also be read out from the 
tables by looking only at the results for 
$s_s^{\downarrow}$ or $s_s^{\uparrow}$ respectively.  

From Tables 1-3, we can calculate the hyperon polarizations
in different cases for the spin states of $s_s$ and $\bar s_s$.
The results are shown in Table 4. 
We see that the produced $\Sigma^0$, $\Sigma^+$ and $n$
are all polarized significantly.
We also see that, similar to $\Lambda$ in 
$e^-p\to e^-K^+\Lambda$, the polarizations 
for $n$ in $e^-p\to e^-\pi^+n$ 
obtained in the three cases 
differ quite significantly from each other. 
But the differences between the results 
for $\Sigma^0$, $\Sigma^+$ are not so significant. 
Hence, to distinguish between the three cases, 
we should study $P_\Lambda$ and $P_n$ in 
the corresponding reactions. 
We note that although the polarization of $\Sigma^0$ 
cannot be measured but it implies a polarization of $\Lambda$,
$P_\Lambda=-P_{\Sigma^0}/3$  
in $e^-p\to e^-\Sigma^0K^+\to e^-\Lambda \gamma K^+$
which can be measured experimentally.

\section{Influences from the sea and the mass effects}

We should emphasize that the results in 
Table 4  are obtained under 
the following two assumptions and/or approximations: 
(1) only valence quarks are taken into account 
in the scattering of $e^-$ with $p$ while 
the sea contribution is neglected;
(2) the influence of the mass differences between 
the $J^P=(1/2)^+$ and the $J^P=(3/2)^+$ baryons 
on the relative production rates are neglected. 
In practice, both of them should be taken into account.
We now discuss their influences on the polarization results 
obtained above. 

It is clear that, in deeply inelastic $e^-p$-scattering, 
the exchanged virtual photon $\gamma^*$ 
can also be absorbed by a sea-quark 
or an anti-sea-quark. 
However, since we consider only the 
exclusive reactions $e^-p\to e^-HK$ 
or the $2\to 2$ process $\gamma^*p\to KH$, 
both of produced hadrons have very 
large momentum fractions 
in the center of mass frame of the hadronic system. 
The sea contributions to such process 
should be very small. 
To see this, we 
take $e^-p\to e^-\Lambda K^+$ as an explicit example.
Besides the case shown in Fig.1, 
this process can also happen 
via the absorption of the $\gamma^*$ by 
a $\bar s_s$ from the sea of the proton.
After the absorption, 
the $\bar s_s$ flies in the direction 
of the $\gamma^*$  while the rest of the 
proton moves in the opposite direction. 
This $\bar s_s$ may pick up a valence 
$u_v$ from the proton and combine into a $K^+$ 
which flies in the direction of the $\bar s_s$ 
while the rest of the proton combines together 
into a $\Lambda$. 
Since the valence quark $u_v$ usually takes a relatively large 
fraction of the momentum of the proton, 
the probability for this to happen 
should be small in particular at high energies. 
Furthermore, 
the ratio of the probability for the virtual photon to 
be absorbed by $\bar s_s$ to that by $u_v$ is given by 
$s(x_B,Q^2)/[4u_v(x_B,Q^2)]$. 
At, e.g., the CEBAF energies, 
it is only of the order of $10^{-3}$.   
Hence we expect that the sea contribution should be negligible. 

In contrast, there should be some influences from 
the mass differences of the 
$J^P=(1/2)^+$ and the $J^P=(3/2)^+$ baryons.
The relative production weights have to be influenced by
such mass differences.
However, it is very interesting to note that in case (a), 
there is completely no influence from the mass
difference on the hyperon polarizations in all the reactions 
discussed above and the $\Lambda$ polarizations
in all the three cases (a), (b) and (c) 
are not influenced by the mass effect. 
This is because in case (a), only $(q_vq_v)_{1,0}q_s^{\downarrow}$
contributes to the production 
of $\Sigma$ or $n$. 
It always gives rise to $\Sigma^{0\downarrow}$,
$\Sigma^{+\downarrow}$ or $n^{\downarrow}$, 
i.e. completely negatively polarized $\Sigma$ or $n$. 
The mass effect
can affect the production rate, but does not influence the
polarization in the corresponding exclusive reaction. 
For $\Lambda$ production, 
only $(ud)_{0,0}$ contributes to $\Lambda$ production
and it contributes only to $\Lambda$ production,
so the $\Lambda$ polarization in all cases are not influenced by
such mass effect.

In cases (b) and (c), there are 
indeed some influences from the mass difference
on the polarizations of $\Sigma^{0}$, $\Sigma^{+}$ and $n$. 
To see how large they may influence 
the polarization of the hyperons, 
we multiply the production rate by 
a corresponding exponential factor $\exp(-\lambda m^2)$
for the production of each baryon, 
where $m$ is the mass of the hyperon. 
We assume that the production processes can be treated as two steps, 
i.e., first the production of the corresponding quarks states then 
the evolution to the corresponding baryons. 
In this way, we need to normalize the production weights  
of the produced baryons to the total production weight 
of the corresponding quark states. 
We obtain that the above polarization 
should be modified into the following.
In case (b), we have
\begin{equation}
P^{(b)}_{\Sigma^+}=-P_eD(y)\cdot 
{2e^{-\lambda m^2_{\Sigma}}+7e^{-\lambda m^2_{\Sigma^*}}\over
6e^{-\lambda m^2_{\Sigma}}+9e^{-\lambda m^2_{\Sigma^*}}}.
\end{equation}
\begin{equation}
P^{(b)}_{\Sigma^0}=-P_eD(y)\cdot 
{2e^{-\lambda m^2_{\Sigma}}+7e^{-\lambda m^2_{\Sigma^*}}\over
6e^{-\lambda m^2_{\Sigma}}+9e^{-\lambda m^2_{\Sigma^*}}}.
\end{equation}
\begin{equation}
P^{(b)}_{n}=P_eD(y)\cdot 
{2e^{-2\lambda m^2_{n}}+
 36e^{-2\lambda m^2_{\Delta}}+19e^{-\lambda (m_n^2+m^2_{\Delta})}\over
 3e^{-2\lambda m^2_{n}}+
 36e^{-2\lambda m^2_{\Delta}}+27e^{-\lambda (m_n^2+m^2_{\Delta})}}.
\end{equation}
Similarly, in case (c), we have
\begin{equation}
P^{(c)}_{\Sigma^+}=-P_eD(y)\cdot
{2e^{-\lambda m^2_{\Sigma}}+4e^{-\lambda m^2_{\Sigma^*}}\over
4e^{-\lambda m^2_{\Sigma}}+5e^{-\lambda m^2_{\Sigma^*}}}.
\end{equation}  
\begin{equation}
P^{(c)}_{\Sigma^0}=-P_eD(y)\cdot
{2e^{-\lambda m^2_{\Sigma}}+4e^{-\lambda m^2_{\Sigma^*}}\over
4e^{-\lambda m^2_{\Sigma}}+5e^{-\lambda m^2_{\Sigma^*}}}.
\end{equation} 
\begin{equation}
P^{(c)}_{n}=-P_eD(y)\cdot
{e^{-2\lambda m^2_{n}}+
 8e^{-\lambda (m_n^2+m^2_{\Delta})}\over
 11e^{-2\lambda m^2_{n}}+
 144e^{-2\lambda m^2_{\Delta}}+100e^{-\lambda (m_n^2+m^2_{\Delta})}}.
\end{equation} 
To get a feeling of how strongly they depend on 
$\lambda$, we show the results in case (b) 
as functions of $\lambda$ in Fig.2.
We see that there are indeed some influences 
on $\Sigma$ or $n$ polarization 
in the corresponding reactions 
but the influences are not very large. 

The following point should be noted here.
All the polarizations obtained above are functions of 
only one variable $y$. 
They are independent of the energies 
under the condition 
that the energies are high enough that the 
parton picture can be used. 

\section{Summary}

In summary, 
we have calculated the polarizations of the baryons in the 
exclusive processes $e^-p\to e^-\Lambda K^+$, 
$e^-p\to e^-\Sigma^+ K^0$, $e^-p\to e^-\Sigma^0 K^+$,
and $e^-p\to e^-n \pi^+$ 
with longitudinally polarized electron beams. 
We used the picture proposed in [\ref{Letter97}],
where it is assumed that the $q_s$ and $\bar q_s$ needed in these
processes to combine with the valence-diquark and struck quark
to form the hyperons and the associated produced meson
have the opposite spins.
The results show that all these baryons are 
longitudinally polarized and the polarizations 
are functions of only one variable $y$.
We compared the results with those 
obtained in other possible cases for 
the spin states of $q_s$ and $\bar q_s$.
We found out that 
the magnitudes of hyperon polarizations 
are considerably large in all the different cases
and that they are quite different 
from each other. 
Hence, they can be used as a good probe 
to study the spin correlations between 
the $q_s$ and $\bar q_s$ in future experiments. 

We thank Li Shi-yuan, Xie Qu-bing and other members of the 
theoretical particle physics group in Shandong University for 
helpful discussions. This work was supported in part by
the National Science Foundation of China (NSFC) and 
the Education Ministry of China under Huo Ying-dong Foundation.

\newpage
\widetext\onecolumn
\begin{table}
\caption{Possible states for the produced $(uus)$,
their relative production weights,
the possible corresponding products and their weights in the reaction    
$e^-p\to e^- \Sigma^+({\rm or\ }\Sigma^{*+})K^0$
in the case that the spins of $s_s$ and ${\bar s_s}$
have same probability to be parallel or anti-parallel 
if the scattered $d_v$ is in the helicity ``$+$'' state.}
\begin{tabular}{lccccccc} \hline
\rule[-0.4cm]{0cm}{1cm}
Possible spin states for $(u_vu_v)s_s$ &
\multicolumn{2}{c}{$(u_vu_v)_{1,0}s_s^\uparrow$} &
\multicolumn{2}{c}{$(u_vu_v)_{1,-1}s_s^\uparrow$} &
\multicolumn{2}{c}{$(u_vu_v)_{1,0}s_s^\downarrow$} &
{$(u_vu_v)_{1,-1}s_s^\downarrow$} \\ \hline
\rule[-0.4cm]{0cm}{1cm}
Corresponding relative weights &
 \multicolumn{2}{c}{$1/6$} &
\multicolumn{2}{c}{$1/3$} &
 \multicolumn{2}{c}{$1/6$} &
{$1/3$} \\ \hline
\rule[-0.4cm]{0cm}{1cm}
Possible products &
 \phantom{s} $\Sigma^{+\uparrow }$    \phantom{s} &
 \phantom{s} $\Sigma^{*+\uparrow }$   \phantom{s} &
 \phantom{s} $\Sigma^{+\downarrow }$  \phantom{s} &
 \phantom{s} $\Sigma^{*+\downarrow }$ \phantom{s} &
 \phantom{s} $\Sigma^{+\downarrow }$    \phantom{s} &
 \phantom{s} $\Sigma^{*+\downarrow }$   \phantom{s} &
\phantom{s} $\Sigma^{*+\Downarrow }$ \phantom{s}  \\ \hline
\rule[-0.4cm]{0cm}{1cm}
Corresponding relative weights &1/3&2/3&2/3&1/3  &1/3&2/3&1 \\ \hline
\rule[-0.4cm]{0cm}{1cm}              
The final relative weights &
  1/18&1/9&2/9&1/9 &1/18&1/9&1/3  \\ \hline
\end{tabular}
\end{table}

\begin{table}
\caption{Possible states for the produced $(uds)$,
their relative production weights,
the possible corresponding products and their weights
in the reaction
$e^-p\to e^-\Lambda({\rm or\ }\Sigma^0 {\rm or\ }\Sigma^{*0})$ $K^+$
in the case that the spins of $s_s$ and ${\bar s_s}$ 
have same probability to be parallel or anti-parallel if 
the scattered $u_v$ is in the helicity ``$+$'' state.}
\begin{tabular}{lccccccccc}\hline 
\rule[-0.4cm]{0cm}{1cm}
Possible spin states & 
 {$(u_vd_v)_{0,0}s_s^\uparrow$} &
\multicolumn{2}{c}{$(u_vd_v)_{1,0}s_s^\uparrow$} &
\multicolumn{2}{c}{$(u_vd_v)_{1,-1}s_s^\uparrow$} &
 {$(u_vd_v)_{0,0}s_s^\downarrow$}&
 \multicolumn{2}{c}{$(u_vd_v)_{1,0}s_s^\downarrow$} &
 {$(u_vd_v)_{1,-1}s_s^\downarrow$} \\ \hline
\rule[-0.4cm]{0cm}{1cm}
Their weights &
 {$3/8$} &
 \multicolumn{2}{c}{$1/24$} &
 \multicolumn{2}{c}{$1/12$}&
 {$3/8$} &
 \multicolumn{2}{c}{$1/24$} &
  {$1/12$} \\ \hline
\rule[-0.4cm]{0cm}{1cm}           
Possible products &
 \phantom{s} $\Lambda ^\uparrow $   \phantom{i} &
 \phantom{s} $\Sigma^{0\uparrow }$  \phantom{} &
 \phantom{} $\Sigma^{*0\uparrow }$ \phantom{} &
 \phantom{s} $\Sigma^{0\downarrow }$    \phantom{} &
 \phantom{} $\Sigma^{*0\downarrow }$   \phantom{} &
 \phantom{s} $\Lambda ^\downarrow $   \phantom{i} &
 \phantom{s} $\Sigma^{0\downarrow }$  \phantom{} &
 \phantom{} $\Sigma^{*0\downarrow }$ \phantom{} &
 \phantom{i} $\Sigma^{*0\Downarrow }$   \phantom{}\\ \hline
\rule[-0.4cm]{0cm}{1cm}
Relative weights
&1&1/3&2/3&2/3&1/3&1 &1/3&2/3&1 \\ \hline
\rule[-0.4cm]{0cm}{1cm}
Final weights &
  3/8&1/72&1/36&1/18 &1/36&3/8&1/72&1/36&1/12  \\ \hline
\end{tabular}
\end{table}

\begin{table}
\caption{Possible states for the produced $(udd)$,
their relative production weights,
the possible corresponding products and their weights in the reaction
$e^-p\to e^-n({\rm or\ }\Delta^0)\pi^+$ 
in the case that the spins of $d_s$ and ${\bar d_s}$
have the same probability to be parallel or anti-parallel 
if the scattered $u_v$ is in the helicity ``$+$'' state.}
\begin{tabular}{lccccccccc}\hline
\rule[-0.4cm]{0cm}{1cm}
Possible spin states  &
 {$(u_vd_v)_{0,0}d_s^\uparrow$} &
 \multicolumn{2}{c}{$(u_vd_v)_{1,0}d_s^\uparrow$} &
 \multicolumn{2}{c}{$(u_vd_v)_{1,-1}d_s^\uparrow$} &
 {$(u_vd_v)_{0,0}d_s^\downarrow$}&
 \multicolumn{2}{c}{$(u_vd_v)_{1,0}d_s^\downarrow$} &
 {$(u_vd_v)_{1,-1}d_s^\downarrow$} \\ \hline
\rule[-0.4cm]{0cm}{1cm}
Their weights &
 {$3/8$} &
 \multicolumn{2}{c}{$1/24$} &
\multicolumn{2}{c}{$1/12$} &
{$3/8$} &
 \multicolumn{2}{c}{$1/24$} &
 {$1/12$} \\ \hline
\rule[-0.4cm]{0cm}{1cm}
Possible products &           
 \phantom{s} $n ^\uparrow $   \phantom{i} &
 \phantom{s} $n^{\uparrow }$  \phantom{i} &
 \phantom{i} $\Delta^{0\uparrow }$ \phantom{} &
 \phantom{s} $n^{\downarrow }$  \phantom{i} &
 \phantom{i} $\Delta^{0\downarrow }$ \phantom{} &
 \phantom{s} $n ^\downarrow $   \phantom{i} &
 \phantom{s} $n^{\downarrow }$  \phantom{i} &
 \phantom{i} $\Delta^{0\downarrow }$ \phantom{} &
 \phantom{i} $\Delta^{0\Downarrow }$   \phantom{}\\ \hline
\rule[-0.4cm]{0cm}{1cm}
Relative weights &1  &1/9&8/9&1/3&2/3&1 &1/9&8/9&1 \\ \hline
\rule[-0.4cm]{0cm}{1cm}
Final weights  &
  3/8 &1/216&1/27&1/36&1/18&3/8 &1/216&1/27&1/12  \\ \hline
\end{tabular}
\end{table}        

\begin{table}
\caption{Polarizations of $\Lambda$, 
$\Sigma^0$, $\Sigma^+$, and $n$
in the reactions $e^-p\to e^-\Lambda ({\rm or\ }\Sigma^0) K^+$,  
$e^-p\to e^- \Sigma^+K^0$ and 
$e^-p\to e^-n \pi^+$
in the three different cases for the spin states of 
the $q_s$ and $\bar q_s$, 
i.e., case (a): the spins of $q_s$ and $\bar q_s$ are anti-parallel,
case (b): the spins of $q_s$ and $\bar q_s$ are parallel,
case (c): the spins of $q_s$ and $\bar q_s$ have 
equal probability to be anti-parallel or parallel.
} 
\begin{tabular}{lcccc} \hline
\rule[-0.4cm]{0cm}{1cm}
Polarizations & $P_{\Lambda}/P_eD(y)$ &
$P_{\Sigma^0}/P_eD(y)$ & $P_{\Sigma^+}/P_eD(y)$ &
$P_{n}/P_eD(y)$ \\ \hline 
\rule[-0.4cm]{0cm}{1cm}
case (a) &
\phantom{ssss} -1 \phantom{sssss}
 & -1 &-1&-1
\\ \hline 
\rule[-0.4cm]{0cm}{1cm}
case (b) & 1 & -3/5 &-3/5 &19/22
\\ \hline
\rule[-0.4cm]{0cm}{1cm}
case (c) & 0 & -2/3 &-2/3 & -3/85
\\ \hline  
\end{tabular}
\end{table} 

\newpage
\begin{figure}
\psfig{file=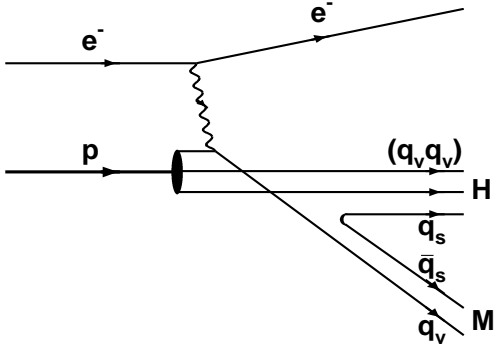,width=8cm}
\caption{Illustrating graph showing the process 
$e^-p\to e^-HK$.} 
\end{figure}

\begin{figure}
\psfig{file=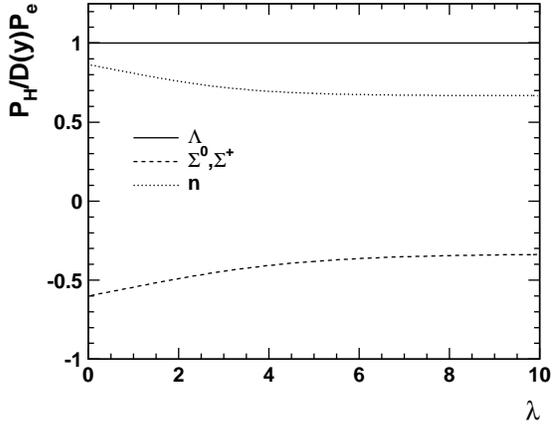,width=9cm}
\caption{Ratio of $P_\Sigma$ or $P_n$ to $P_eD(y)$ 
as a function of $\lambda$ for case (b) 
in different processes from Eqs.(6-8).}
\end{figure}

\newpage

\noindent
Table captions

\vskip 0.3cm\noindent
Table 1: 
Possible states for the produced $(uus)$,
their relative production weights,
the possible corresponding products and their weights in the reaction    
$e^-p\to e^- \Sigma^+({\rm or\ }\Sigma^{*+})K^0$
in the case that the spins of $s_s$ and ${\bar s_s}$
have same probability to be parallel or anti-parallel 
if the scattered $d_v$ is in the helicity ``$+$'' state.

\vskip 0.3cm\noindent
Table 2:
Possible states for the produced $(uds)$,
their relative production weights,
the possible corresponding products and their weights
in the reaction
$e^-p\to e^-\Lambda({\rm or\ }\Sigma^0 {\rm or\ }\Sigma^{*0})$ $K^+$
in the case that the spins of $s_s$ and ${\bar s_s}$ 
have same probability to be parallel or anti-parallel if 
the scattered $u_v$ is in the helicity ``$+$'' state.

\vskip 0.3cm\noindent
Table 3:
Possible states for the produced $(udd)$,
their relative production weights,
the possible corresponding products and their weights in the reaction
$e^-p\to e^-n({\rm or\ }\Delta^0)\pi^+$ 
in the case that the spins of $d_s$ and ${\bar d_s}$
have the same probability to be parallel or anti-parallel 
if the scattered $u_v$ is in the helicity ``$+$'' state.

\vskip 0.3cm\noindent
Table 4:
Polarizations of $\Lambda$, 
$\Sigma^0$, $\Sigma^+$, and $n$
in the reactions $e^-p\to e^-\Lambda ({\rm or\ }\Sigma^0) K^+$,  
$e^-p\to e^- \Sigma^+K^0$ and 
$e^-p\to e^-n \pi^+$
in the three different cases for the spin states of 
the $q_s$ and $\bar q_s$, 
i.e., case (a): the spins of $q_s$ and $\bar q_s$ are anti-parallel,
case (b): the spins of $q_s$ and $\bar q_s$ are parallel,
case (c): the spins of $q_s$ and $\bar q_s$ have 
equal probability to be anti-parallel or parallel.

\vskip 1cm\noindent
Figure captions

\vskip 0.3cm\noindent
Figure 1: 
Illustrating graph showing the process $e^-p\to e^-HK$.

\vskip 0.3cm\noindent
Figure 2: 
Ratio of $P_\Sigma$ or $P_n$ to $P_eD(y)$ 
as a function of $\lambda$ for case (b) 
in different processes from Eqs.(6-8).

\end{document}